# Isotropic and anisotropic heat transfer in active wall porous media foam type

DEPTULSKI Rafael C. [a], BENNACER Rachid [a*]

[a] LMT-Cachan / ENS Paris-Saclay / CNRS / Université Paris-Saclay, 61 avenue du Président Wilson, 94230 Cachan, France

*Corresponding author: *rachid.bennacer@ens-paris-saclay.fr*

**Abstract:** Positive buildings in energy are, nowadays, a recurrent objective of many researches in the construction and energetic efficiency domain. Furthermore, to achieve this objective, some studies about active and reactive walls have been carried out employing porous medium as a main structure. Nevertheless, transfer characterization in a foam type sample is not fully understood. The goal of this study is to improve the characterization of heat transfer in isotropic and anisotropic configurations of a porous medium. Thus, a finite volume method was implemented to study a heat transfer through these media, in the interest of achieving their ratio equivalent to fluid thermal conductivity (*i.e.* Nusselt number). Finally, the results indicate a notable influence of the ratio of the contact and the total inlet area on the isotropic configuration as well as strong influence given by the different axis on the anisotropic model. Moreover, the analysis shows that in an active wall constituted by two solid phases, these effects will be preponderant for their characterization.

## Introduction

Porous medium foam type is generally considered as an efficient heat exchanger. Following preceding study [1], the coupling between the heat and the mass transfer in an orthogonal reference is suitable to form an active wall. The desirable structure here studied is an active and reactive wall, formed by a continuous solid thermal conductive matrix that provides a structural achievement, then, a second solid phase that coats the skeletal phase to get a perfect insulator in a thermal activated mode and, lastly, a void phase that has an important volumetric fraction in an inactivated configuration.

Many research projects pursue to explain the flow and the heat behaviour in porous media [2] in order to predict their equivalent properties in different compositions such as metallic, polymeric or mixed samples foam type [3–5]. Different applications in this domain are increasingly performed thanks to its high mass and thermal transfer capacity given by an impressive specific surface area and permeability in addition to their significant mechanical resistance [6,7]. Hence, the goal is to explore this thermomechanical potential to construct an active wall in porous media. Therefore, this system has many functions and can be combined to work with several thermo-activate applications such as deformable porous polymers membranes [8] in our case or phase change materials [9] in an energy storage framework.

This work does not seek beyond the study of thermal transfer in isotropic and anisotropic porous media, however, a multiphasic model of an active wall will be present to contextualize the structure. Firstly, the methodology for creating the porous medium in its different configurations is presented, followed by the numerical method of heat transfer used. Finally, the results and conclusions are presented in order to answer the degree of influence of the non-linearity of a porous medium for its heat transport.

# 1. Analysis and modelling

This part covers the generating process of a 3D porous medium framework divided into three points: generating a random porous structure, controlling its morphology in order to create an anisotropic configuration, as well as numerical modelling.

Here, different structures were presented, firstly in two phases (*i.e.* solid-fluid configuration) and then in three phases (*i.e.* solid-solid-fluid configuration). These samples were originally generated in a numerical device, based on correlated Gaussian random fields [10], in a cube with a triple correlation length of 10%, a standard deviation of 1%, an expectation of 0 and 250 eigenmodes retained (Figure 1 (a)). Thus, the designed structure results in a Representative Elementary Volume (REV) of a foam-like porous with 90% voids (Figure 1 (b) and (c)). Hither, the two-phase model is analysed in order to achieve an isotropic and anisotropic transfer characterization, besides that, the three-phase model is presented to illustrate the operation of an active and reactive wall through a porous medium.

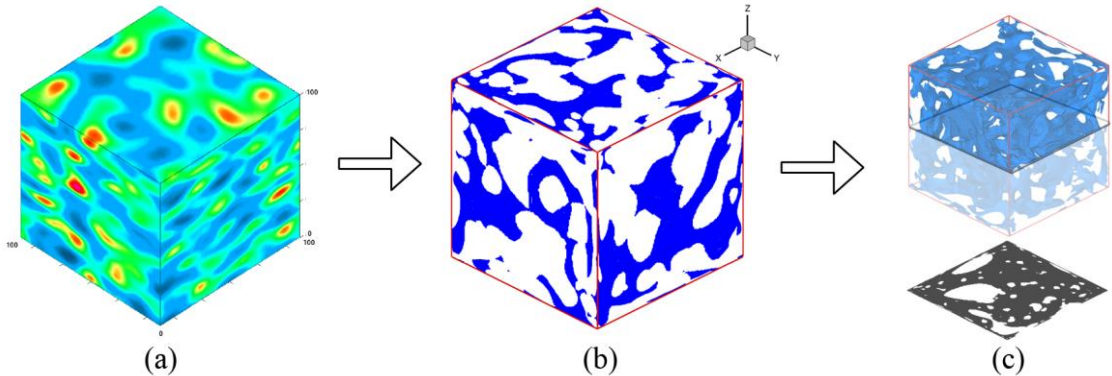

Figure 1: Development stages to generate a porous structure based on Gaussian random draws. Correlated Gaussian random field (a) and 3D structure 90% voids (b) and (c)

Hence, Figure 2 shows the 3D model, at the pore scale, of an active and reactive wall in a considered homogeneous structure by their solid skeletal matrix (Phase 1) and a coat solid phase (Phase 2) that involves this first phase.

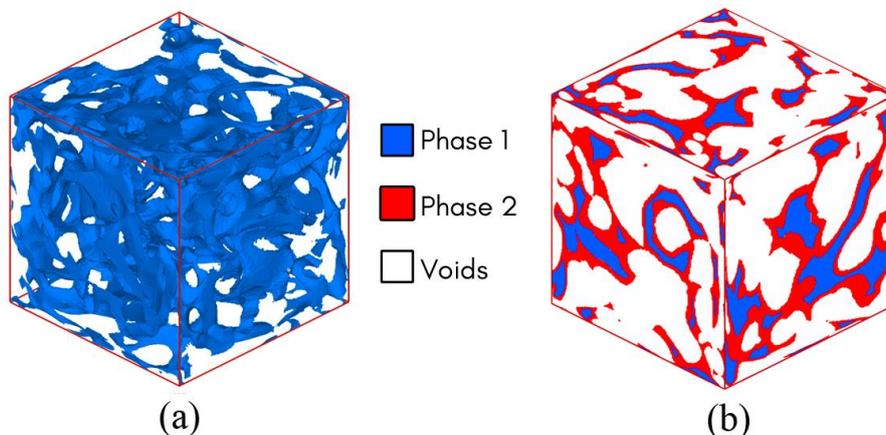

Figure 2: 3D view of an active and reactive wall in an homogeneous porous medium with skeletal and coat phases

This proposed structure is still little studied and consisting of a rigid solid matrix which takes the heat energy to the second non-rigid solid matrix and, therefore, result in a system of expansion-shrinkage which gives to this couple an inhomogeneous nature in order to get a perfect insulator. In this case, after an initial request, an inhomogeneous structure will be formed and its response to external requests will no longer have the expected standard nature. Consequently, in order to understand this new heat transfer behaviour, an inhomogeneous structure has been generated based on the first homogeneous structure (Figure 2) by applying a dilatation threshold gradient from -30% to +30% at the two opposite sides in the vertical direction, as shown in Figure 3 and Figure 4.

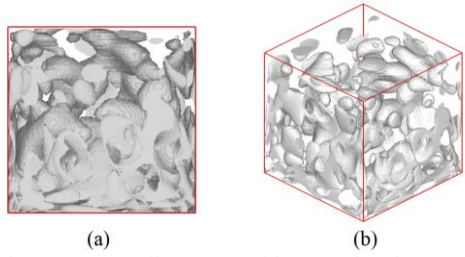

Figure 3: 3D slice (a) and isometric view (b) of an inhomogeneous porous medium by applying a dilatation gradient from -30% to +30% in vertical axis

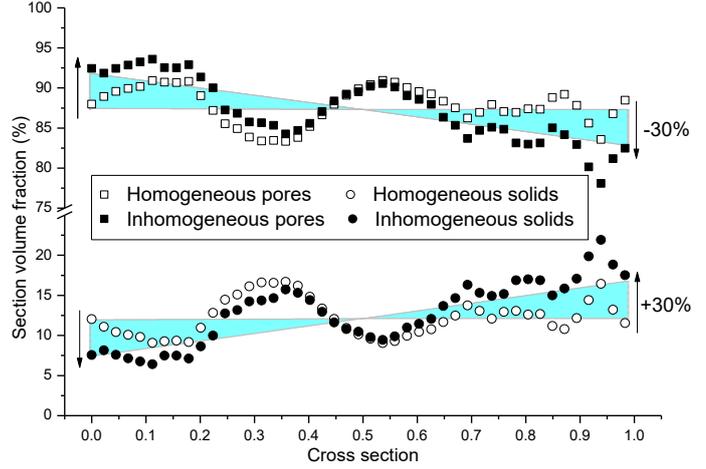

Figure 4: Volume fraction distribution of the phases along the volume for the two configurations

To analyse these different structures and understand their behaviours, a finite volume method is implemented, as the presented scheme in Figure 5, in order to discretize a heat flux through porous media and it has been performed solving the momentum and energy equations.

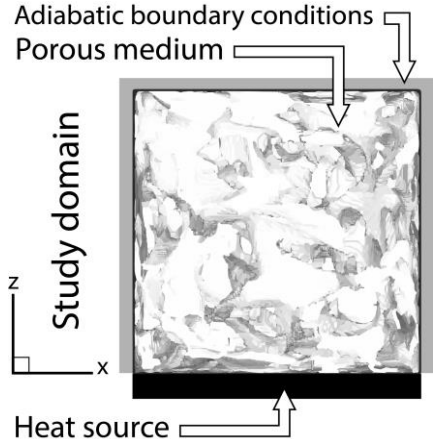

Figure 5: Porous media heat exchanger configuration

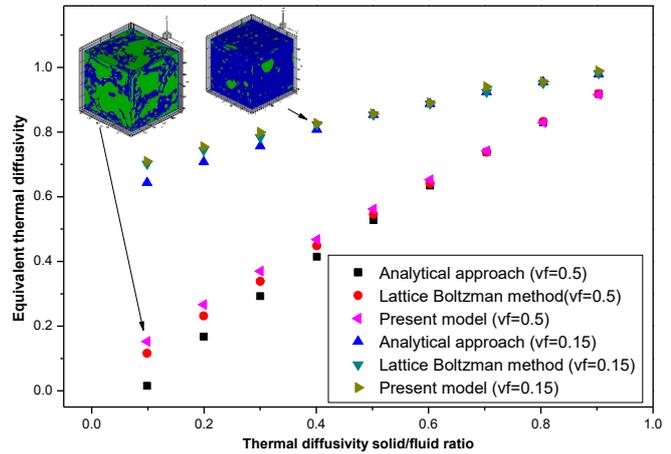

Figure 6: Numerical method validation

Continuity equation:

$$\nabla \vec{V} = 0 \quad (1)$$

Momentum equation:

$$\frac{\partial \vec{V}}{\partial t} + (\vec{V} \cdot \nabla)\vec{V} = -\nabla P + \frac{1}{Re}\nabla(\mu^* \nabla \vec{V}) + \frac{1}{Da}\vec{V} + \frac{Ra}{Pr\, Re^2}(\theta + NS)\vec{k} \quad (2)$$

Energy equation:

$$\frac{\partial \theta}{\partial t} + \vec{V} \cdot \nabla \theta = \frac{1}{Pr}\nabla(\lambda^* \nabla \theta) \quad (3)$$

The continuity of both temperatures the heat flux on the solid-fluid interface is employed:

$$\theta_s = \theta_f \quad \therefore \quad \frac{\partial \theta_f}{\partial n} = \lambda^* \frac{\partial \theta_s}{\partial n} \quad (4)$$

To reach a stable state solution, a convergence criterion based on the temperature field is adopted. This norm is defined by the following conditions:

$$\frac{\sum_{i,j}|\varphi_{i,j}^{m+1} - \varphi_{i,j}^m|}{\sum_{i,j}|\varphi_{i,j}^{m+1}|} \leq \epsilon = 10^{-5} \quad (5)$$

Where $m$ is an interaction counter and $\epsilon$ is the relative error. Furthermore, some validation tests have been performed in Figure 6 and this method has been presented good accuracy conforming to reference results published in several preceding works [7–9], as well as qualitatively experimental results validation [14] and instability verification [15]. Moreover, the criterion of validity concerning this method is considered by $0.1 \leq \lambda^* \leq 200$ and $Ra = 0$.

## 2. Results and discussion

In a homogeneous foam type configuration, heat transfer is significantly affected by the continuous nature of the skeletal phase. Moreover, here we analyse the isotropic nature of this medium and the influence of the ratio of the contact and the total inlet area $R$ on the Nusselt number (*i.e.* $\lambda_{eq}/\lambda_{fluid}$ in this porous configuration) after heating the sample by the three different axis x, y and z (Figure 7). Hither, this section search to achieve the isotropic nature of heat transfer in the study domain.

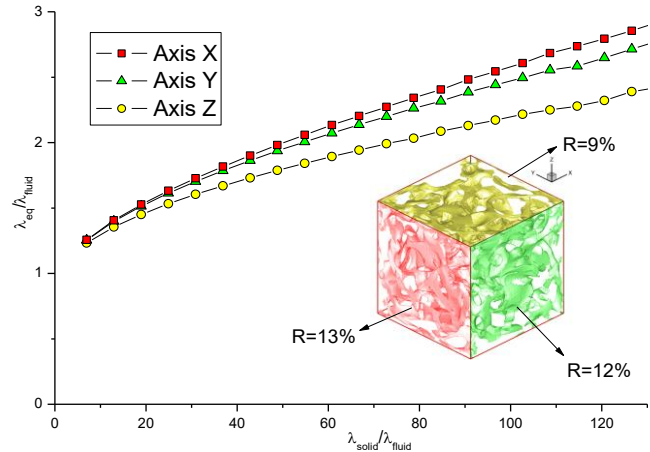

Figure 7: Nusselt number for the three-different axis x, y and z and their respective contact surface areas in an isotropic structure

Figure 7 shows some small variations between the different solicitations axis according to the increase of $\lambda^*$, despite that, this structure is still considered as isotropic. Thus, according to the same figure, it is evident that the heat transfer in an isotropic model varies according to its ratio of the contact and the total inlet area $R$ (*i.e.* 13%, 12% and 9% to x, y, and z respectively). Such connection between the material and the solid surface is a key point equivalent (in opposing way) to the thermal resistance effect. Besides that, we observe the heat transfer behaviour in an inhomogeneous configuration (Figure 3) following their Nusselt number after heating the sample by the three different axis x, y and z (Figure 8).

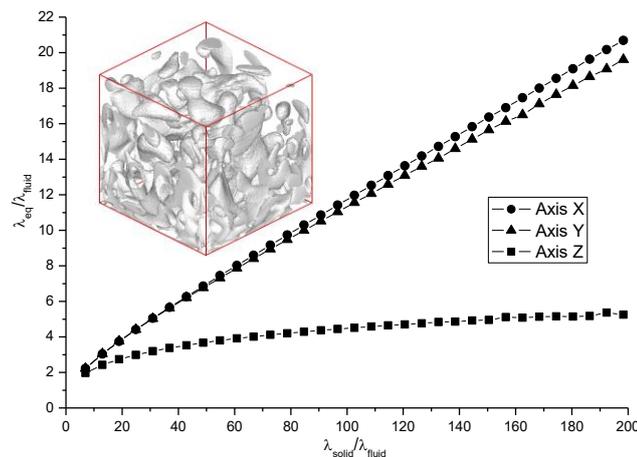

Figure 8: Nusselt number for the three-different axis x, y and z in an anisotropic structure

In this case, the result shows that on the x and y-axis, a continuous solid matrix is present and connecting the opposed side, that provided an important heat conduction. On the other hand, on the z-axis, there is no more direct connection (*i.e.* absence of percolation effect) between their opposite sides, consequently, the heat transfer is considerably less important. These analyses prove that this structure presents an anisotropic heat transfer vector as well as its morphology.

# Conclusions

This review explored the heat transfer behaviour in a porous medium subjected to both morphology configurations: isotropic and anisotropic. Then, using a stochastic generation of 3D structures based on correlated random fields, a foam-like porous medium was constructed with a continuous skeletal solid matrix filled with non-regular voids for the purpose of illustrating the operating role of an active wall.

Thus, a finite volume discretization method has been performed in order to characterize the thermal conductivity in a non-linear system. The Nusselt number, for the proposed 90% voids isotropic model, increases according to the ratio of the contact and the total inlet area of which solicitation axis. In addition, the results of the anisotropic configuration showed that if the solid matrix is not continuous between inlet and outlet sides the equivalent thermal properties are significantly affected.

Lastly, understanding the issue about the isotropic and anisotropic thermal behaviour is a preponderant step towards the construction of an active and reactive wall. This development will be useful to open up new research possibilities and to further investigate the abilities to this structure manages its transfer properties by thermo-activation.

## Nomenclature

| | | | |
|---|---|---|---|
| $k$ | orthogonal vector | $\varphi$ | heat flux density, $W.m^{-2}$ |
| $Nu$ | Nusselt number | $\phi$ | heat flux, $W$ |
| $Pr$ | Prandtl number | $\varepsilon$ | porosity, % |
| $R$ | ratio of the contact and the total area, % | $\theta$ | dimensionless temperature |
| $Ra$ | Rayleigh number | $\lambda$ | thermal conductivity, $W.m^{-1}.K^{-1}$ |
| $Re$ | Reynolds number | **Subscripts / Superscripts** | |
| $Sc$ | Schmidt number | $eq$ | equivalent value |
| $t$ | dimensionless time | $f$ | fluid phase |
| $V$ | dimensionless velocity | $n$ | normal vector |
| **Greek** | | $s$ | solid phase |
| $\mu$ | dynamic viscosity, $Pa.s$ | $*$ | solid to fluid ratio |